\documentclass[12pt]{article}
\usepackage{a4wide,latexsym,graphicx,epsfig,psfrag,here}
 
\newcommand{\ol}{\overline}

\newcommand{\epsp}{\epsilon^\prime}

\newcommand{\bea}{\begin{eqnarray}}
\newcommand{\eea}{\end{eqnarray}}
\newcommand{\beq}{\begin{equation}}
\newcommand{\eeq}{\end{equation}}
\newcommand{\nn}{\nonumber}
\newcommand{\nl}{\nonumber\\}

\newcommand{\ra}{\rightarrow}

\newcommand{\PL}[3]{{Phys. Lett.} {\bf#1} {(#2)} {#3}} 
\newcommand{\PRL}[3]{{Phys. Rev. Lett.}  {\bf#1} {(#2)} {#3}} 
\newcommand{\PR}[3]{{Phys. Rev.} {\bf#1} {(#2)} {#3}} 
\newcommand{\NP}[3]{{Nucl. Phys.} {\bf#1} {(#2)} {#3}} 
\newcommand{\EPJ}[3]{{Eur. Phys. J.} {\bf#1} {(#2)} {#3}} 
\newcommand{\ZP}[3]{{Z. Phys.} {\bf#1} {(#2)} {#3}} 
\newcommand{\hepph}[1]{{\tt hep-ph/#1}}

\newcommand{\cL}{{\cal L}}
\newcommand{\cM}{{\cal M}}

\newcommand{\ba}{\begin{array}{c}}
\newcommand{\bat}{\begin{array}{cc}}
\newcommand{\ea}{\end{array}}

\newcommand{\mbf}{\mathbf} 
 
\def\slashchar#1{\setbox0=\hbox{$#1$}\dimen0=\wd0%
\setbox1=\hbox{/}\dimen1=\wd1%
\ifdim\dimen0>\dimen1%
\rlap{\hbox to
\dimen0{\hfil/\hfil}}#1\else                                     
\rlap{\hbox to \dimen1{\hfil$#1$\hfil}}/\fi}
\newcommand{\wh}{\widehat}

\normalsize 
\begin{document} 
\begin{titlepage} 
\begin{flushright}
IFIC/03-23 \\
UWThPh-2003-8\\ 
May 2003\\ 
\end{flushright} 
\vspace{2.5cm} 
\begin{center} 
{\Large \bf Meson resonances, large $N_c$ and chiral symmetry} \\[40pt]

V. Cirigliano$^{1}$, G. Ecker$^{2}$, H. Neufeld$^{2}$ 
and A. Pich$^{1}$ 
 
\vspace{1cm}
${}^{1)}$ Departament de F\'{\i}sica Te\`orica, IFIC, CSIC --- 
Universitat de Val\`encia \\ 
Edifici d'Instituts de Paterna, Apt. Correus 22085, E-46071 
Val\`encia, Spain \\[10pt]
 
${}^{2)}$ Institut f\"ur Theoretische Physik, Universit\"at 
Wien\\ Boltzmanngasse 5, A-1090 Vienna, Austria \\[10pt] 
\end{center} 
 
\vfill 

\begin{abstract}\noindent
We investigate the implications of large $N_c$ and chiral symmetry for
the mass spectra of meson resonances. Unlike for most other mesons, the 
mass matrix of the light scalars deviates strongly from its
large-$N_c$ limit. We discuss the possible assignments for the
lightest scalar nonet that survives in the large-$N_c$ limit.
\end{abstract}

\vfill
 
\noindent 
*~Work supported in part by HPRN-CT2002-00311 (EURIDICE) and by
Acciones Integradas, Project No. 19/2003 (Austria),
HU2002-0044 (MCYT, Spain).
\end{titlepage} 
\newpage

\addtocounter{page}{1} 
\paragraph{1.}
The interpretation of scalar meson resonances has been controversial
for a long time. The problems are of both experimental and theoretical
nature \cite{st02,rpp02}. As a distinctive feature of scalar mesons,
the $SU(3)$ singlet has vacuum quantum numbers. Scalar mesons may 
therefore be especially susceptible to the non-trivial structure of
the QCD vacuum.

To explore the peculiar properties of $0^{++}$ mesons,
we propose a general analysis of the mass spectra of all light meson
resonances that is only based on established consequences of QCD
for light hadrons. In particular, we make no reference to the internal
structure of meson resonances ($\ol{q}q$, multi-quark states,
meson-meson bound states, glueballs, \dots). 

Our main assumptions are two-fold.
\begin{enumerate} 
\item[i.] We assume that the mass splittings of light meson
   multiplets and their couplings to pseudoscalar mesons can be understood
   in the framework of a chiral resonance Lagrangian \cite{egpr89,eglpr89}. 
   Only leading terms in the chiral expansion will be considered.
\item[ii.] In first approximation, we assume a nonet structure for the
  mesons as predicted by QCD in the limit of large $N_c$
  \cite{witten}. In order to parametrize the deviations from the nonet
  limit, we include in a second step all possible sub-leading terms in
  $1/N_c$ of relevance for the mass spectrum as long as they are of
  leading order in the chiral expansion.
\end{enumerate} 

\paragraph{2.}
We first recall the main features of chiral resonance theory
\cite{egpr89}. The resonance fields come in $SU(3)$ octets and
singlets  and they transform in
the usual way under a non-linear realization of chiral $SU(3)$. The
octet ($R_i$) and singlet ($R_0$) fields are grouped together
in a nonet field $R$:
\begin{equation} 
R = \lambda_i R_i/\sqrt{2} + R_0/\sqrt{3}~\bf 1 ~.
\end{equation}
In the limit of large $N_c$, these nine fields are degenerate in the
chiral limit with a common mass $M_R$. To understand the
phenomenological values of the low-energy constants (LECs) $L_i$ in
the chiral Lagrangian of $O(p^4)$ \cite{gl85a}, a chiral
resonance Lagrangian of the following generic form is employed:
\begin{eqnarray}  
\cL_R &=& \displaystyle\frac{1}{2} \langle \nabla R \cdot \nabla R -
M_R^2 R^2 \rangle +\langle R g_2^R \rangle ~.
\label{SLagr4}
\end{eqnarray}
Following the notation of Ref.~\cite{egpr89}, $\nabla R$ denotes a
chiral- and gauge-covariant derivative. All space-time indices
are omitted.  $g_2^R$ is a chiral field of $O(p^2)$ that couples to
the respective resonance multiplet of given spin-parity;
$\langle \dots \rangle$ stands for the three-dimensional flavour
trace. The large-$N_c$ relations 
for the scalar couplings discussed in Ref.~\cite{egpr89} are automatically 
reproduced by the Lagrangian (\ref{SLagr4}).

In order to calculate the contributions of meson resonance exchange to
the LECs of $O(p^6)$ \cite{bce99} , the Lagrangian (\ref{SLagr4}) must
be extended:
\begin{eqnarray}  
\cL_R &=& \displaystyle\frac{1}{2} \langle \nabla R \cdot  \nabla R -
M_R^2 R^2 \rangle +\langle R (g_2^R + g_4^R) \rangle + 
\langle R^2 h_2^R + ~\dots ~\rangle ~. 
\label{SLagr6}
\end{eqnarray}
Only single flavour traces appear in (\ref{SLagr6}) because we assume
large $N_c$ at this point. For our purposes, we only need to consider
bilinear interaction terms of the type shown in (\ref{SLagr6})
where $R^2$ (and therefore $h_2^R$) is a
Lorentz scalar. There are other bilinear terms that
will also contribute at $O(p^6)$, e.g., mixed terms with different
resonance fields. On the other hand,
cubic and higher couplings in the resonance
fields do not contribute to the effective low-energy Lagrangian of
$O(p^6)$. The order of the chiral fields $g_i^R, h_i^R$ is indicated by
the subscript. The resonance Lagrangian (\ref{SLagr6}) induces the
following contribution to the effective Lagrangian to $O(p^6)$:
\begin{eqnarray} 
\cL_{\rm eff} &=& \displaystyle\frac{1}{2 M_R^2}\langle g_2^R g_2^R
\rangle \nl
&+& \displaystyle\frac{1}{2 M_R^4}\langle \nabla g_2^R \cdot \nabla
g_2^R \rangle +
\displaystyle\frac{1}{M_R^4}\langle g_2^R g_2^R h_2^R \rangle +
\displaystyle\frac{1}{M_R^2}\langle g_2^R g_4^R \rangle + ~\dots
\label{Seff}
\end{eqnarray} 
The first line reproduces the result of Ref.~\cite{egpr89}. The
second line contains the contributions of $O(p^6)$ from the 
exchange of a specific resonance multiplet with the Lagrangian
(\ref{SLagr6}). 

Here we are interested in the mass splittings of the mesons.
The resonance masses are derived from the non-derivative bilinear part of
the Lagrangian (\ref{SLagr6}):
\begin{eqnarray}  
\cL_R^{\rm mass} &=& - \displaystyle\frac{M_R^2}{2} \langle R^2 \rangle +
e_m^R \langle R^2 \chi_+ \rangle \label{Lmass} \\
h_2^R &=& e_m^R \chi_+ + \dots~, \nn
\end{eqnarray} 
with coupling constant $e_m^R$.
The chiral field $\chi_+$ contains the quark mass matrix
$\cM_q$~: 
\begin{equation} 
\chi_+ = 4 B \cM_q + \dots
\end{equation} 
where $B$ is related to the scalar condensate \cite{gl85a}. We always
stay in the $SU(2)$ limit with $M_\pi^2= B(m_u + m_d)= 2 B \hat{m}$,
$M_K^2 = B(m_s + \hat{m})$.

The structure of the mass Lagrangian (\ref{Lmass}) is the same for all
meson resonances with $R^2$ the appropriate
bilinear field combination \cite{egpr89}. To leading order both in large
$N_c$ and in the chiral expansion, the mass splittings in a nonet are
governed by a single coupling constant $e_m^R$. Of course, this constant
will in general be different for different resonance nonets.

We use the following notation for the various resonance fields and
for the corresponding masses:\\
\begin{tabular}{lrll}
\mbox{         } \hspace*{1.5cm} & $R_{I=1}$ & \mbox{         } &
\hspace*{2cm} isotriplet fields, \\
\mbox{         } & $R_{I=1/2}$ & \mbox{         } & \hspace*{2cm} 
isodoublet fields,\\
\mbox{         } & $R_0, R_8$ & \mbox{         } & \hspace*{2cm}
singlet and isosinglet octet fields,\\
\mbox{         } & $R_H, R_L$ & \mbox{         } & \hspace*{2cm}
isosinglet mass eigenfields.
\end{tabular}
\vspace*{.3cm}

\noindent
The masses of the non-singlet fields can immediately be extracted
from the Lagrangian (\ref{Lmass}):
\begin{eqnarray} 
M_{I=1}^2 &=& M_R^2 - 4 e_m^R M_\pi^2 \nl
M_{I=1/2}^2 &=& M_R^2 - 4 e_m^R M_K^2 ~, \label{nonsinglet}
\end{eqnarray} 
implying
\begin{eqnarray}
e_m^R &=& \displaystyle\frac{M_{I=1}^2 - M_{I=1/2}^2}
{4(M_K^2 - M_\pi^2)} \nl
M_R^2 &=& M_{I=1}^2 + \displaystyle\frac{M_\pi^2(M_{I=1}^2 -
  M_{I=1/2}^2)}{M_K^2 - M_\pi^2} ~.\label{MS}  
\end{eqnarray}

\paragraph{3.} 
The mass matrix $M_0^2$ for the isosinglet fields $R_8, R_0$  is 
obtained from (\ref{Lmass}) as
\begin{equation} 
M_0^2 = \left ( \bat M_R^2 - \displaystyle\frac{4}{3}e_m^R (4 M_K^2 - 
M_\pi^2) &
\displaystyle\frac{8\sqrt{2}}{3}e_m^R (M_K^2 - M_\pi^2) \\
\displaystyle\frac{8\sqrt{2}}{3}e_m^R (M_K^2 - M_\pi^2) &
M_R^2 - \displaystyle\frac{4}{3}e_m^R (2 M_K^2 + M_\pi^2)
\ea \right)~, \label{M0}
\end{equation}
with eigenvalues
\begin{eqnarray} 
M_R^2 - 4 e_m^R M_\pi^2 & \hspace*{1cm} {\rm and} \hspace*{1cm} &
M_R^2 - 4 e_m^R (2 M_K^2 - M_\pi^2)~. 
\end{eqnarray}
In terms of the non-singlet masses (\ref{nonsinglet}),
the isosinglet masses are given by
\begin{eqnarray} 
M_L^2 &=& M_{I=1/2}^2 - | M_{I=1/2}^2 - M_{I=1}^2 | \nl
M_H^2 &=& M_{I=1/2}^2 + | M_{I=1/2}^2 - M_{I=1}^2 | ~.
\label{singlet}
\end{eqnarray}
The mass matrix (\ref{M0})
can be diagonalized with an orthogonal matrix $O$:
\begin{equation} 
M_0^2 = O^T M_D^2 O ~, \qquad M_D^2={\rm diag}(M_H^2,M_L^2)
\end{equation}
$$
O = \left( \bat \cos{\theta} & - \sin{\theta} \\
                \sin{\theta} & \cos{\theta} \ea \right)~,
$$
where the mixing angle $\theta$ is defined mod $\pi$. It will be
convenient to consider the interval $- \pi/2 \le \theta \le \pi/2$.
The mass eigenfields $R_H, R_L$ are then given by
\begin{eqnarray} 
R_H &=& \cos{\theta} ~R_8 - \sin{\theta} ~R_0 \nl
R_L &=& \sin{\theta} ~R_8 + \cos{\theta} ~R_0 ~.
\end{eqnarray} 

The fields $R_8, R_0$ can be expressed in terms of fields with
specific flavour content in the $\ol{q}q$ picture:
\begin{eqnarray}  
R_8 &=& \displaystyle\frac{1}{\sqrt{3}} R_{\rm non-strange} -
\sqrt{\frac{2}{3}} R_{\rm strange} \nl
R_0 &=& \sqrt{\frac{2}{3}} R_{\rm non-strange} +
\displaystyle\frac{1}{\sqrt{3}} R_{\rm strange}~.
\label{flavour}
\end{eqnarray} 
The mass eigenfields can then also be written as
\begin{eqnarray} 
R_H &=& \displaystyle\frac{1}{\sqrt{3}}(\cos{\theta}-
\sqrt{2}\sin{\theta}) R_{\rm non-strange}-
\displaystyle\frac{1}{\sqrt{3}}(\sqrt{2}\cos{\theta}+
\sin{\theta}) R_{\rm strange} \nl
R_L &=& \displaystyle\frac{1}{\sqrt{3}}(\sqrt{2}\cos{\theta}+
\sin{\theta}) R_{\rm non-strange}+
\displaystyle\frac{1}{\sqrt{3}}(\cos{\theta}-
\sqrt{2}\sin{\theta}) R_{\rm strange}  ~.
\label{SHSL}
\end{eqnarray}  
Ideal mixing with $R_H = - R_{\rm strange}$, $R_L = R_{\rm
 non-strange}$ corresponds to
\begin{equation} 
\tan{\theta_{\rm ideal}}=1/\sqrt{2} \quad \ra \quad \theta_{\rm ideal} 
= 35.3^\circ ~.
\end{equation}

The mass matrix (\ref{M0}) has a special property as already noted in
Ref.~\cite{black99}: the mixing angle
$\theta$ depends only on the sign but not on the magnitude of
$e_m^R$. We now discuss the two possibilities in turn.\\[.4cm] 
{\bf i.}~~$\mathbf{e_m^R > 0}$ \\[.2cm] 
The resonance masses are ordered as 
\begin{eqnarray} 
& M_L < M_{I=1/2} < M_H = M_{I=1}~.& \label{order+}
\end{eqnarray} 
The mixing angle is found to be
\begin{eqnarray}
\theta_{e_m^R >0} = - \arctan{\sqrt{2}}
\quad \ra \quad \theta_{e_m^R >0}= -54.7^\circ ~.
\end{eqnarray} 
The ordering of masses is unusual because the strange member of the
octet has a smaller mass than the isotriplet state. 
Most resonance nonets do not display such an inverted hierarchy.
Also the mixing pattern is unusual (dual ideal mixing 
\cite{black99}): the light neutral
field $R_L$ is identical to $R_{\rm strange}$ and
$R_H=R_{\rm non-strange}$.\\[.4cm] 
{\bf ii.}~~$\mathbf{e_m^R < 0}$ \\[.2cm] 
The isotriplet now changes position compared to (\ref{order+}):
\begin{eqnarray} 
&  M_{I=1} = M_L < M_{I=1/2} < M_H  ~.& \label{order-}
\end{eqnarray} 
The mixing angle is now
\begin{eqnarray}
\theta_{e_m^R <0} = \arctan{\displaystyle\frac{1}{\sqrt{2}}}
\quad \ra \quad \theta_{e_m^R <0}= \theta_{\rm ideal}= 35.3^\circ~, 
\end{eqnarray}
and therefore
\begin{eqnarray} 
R_L = R_{\rm non-strange}~, \qquad & \qquad  R_H = - R_{\rm strange}
~. \label{nonet}
\end{eqnarray}

This pattern is very well satisfied by the vector mesons and, 
as we shall review below, at least approximately also by other nonets:
\begin{eqnarray} 
& M_\rho \simeq M_\omega < M_{K^*} < M_\phi \nl
& 2 M_{K^*}^2 \simeq M_\omega^2 +  M_\phi^2 \\
& \theta \simeq \theta_{\rm ideal} ~. \nn
\end{eqnarray}
Therefore, only the case $e_m^R < 0$ corresponds to the usual notion of
a nonet with ideal mixing \cite{okubo63}.

\paragraph{4.}
Of course, not even the vector mesons are ideally mixed. 
We consider here a minimal version of nonet symmetry breaking where
only the terms bilinear in the resonance fields are affected.

To lowest order in the chiral expansion, the mass Lagrangian
(\ref{Lmass}) acquires two additional terms that are sub-leading in
$1/N_c$:
\begin{equation} 
\cL_R^{\rm mass} = - \displaystyle\frac{M_R^2}{2} \langle R^2 \rangle +
e_m^R \langle R^2 \chi_+ \rangle + k_m^R R_0 \langle \wh{R} \chi_+ \rangle 
- \displaystyle\frac{\gamma_R M_R^2}{2} R_0^2~, 
\label{Lnl}
\end{equation} 
where  $\wh{R}=\lambda_i R_i/\sqrt{2}$ is the octet field. 
The Lagrangian (\ref{Lnl}) is the most general
lowest-order chiral Lagrangian bilinear in octet and singlet fields
that can contribute to the mass matrix. All other terms can be
absorbed by a redefinition of the parameters in (\ref{Lnl}).

The non-singlet fields $R_{I=1}$, $R_{I=1/2}$ are unaffected and their
masses are still given by (\ref{nonsinglet}). The type of hierarchy is
again determined by the sign of $e_m^R$. The additional parameters  
$k_m^R$ and $\gamma_R$ give rise to the isosinglet mass matrix
\begin{equation} 
M_0^2 = \left ( \bat M_R^2 - \displaystyle\frac{4}{3}e_m^R (4 M_K^2 - 
M_\pi^2) &
\displaystyle\frac{8\sqrt{2}}{3}(e_m^R + \displaystyle\frac{\sqrt{3}}{2} 
k_m^R) (M_K^2 - M_\pi^2) \\
\displaystyle\frac{8\sqrt{2}}{3}(e_m^R + \displaystyle\frac{\sqrt{3}}{2} 
k_m^R) (M_K^2 - M_\pi^2)
&
M_R^2(1 + \gamma_R) - \displaystyle\frac{4}{3}e_m^R (2 M_K^2 + M_\pi^2)
\ea \right)~. \label{M0corr}
\end{equation}
The interpretation of this mass matrix is straightforward.
The first entry corresponds to the isosinglet octet field $R_8$ and 
it satisfies a (quadratic) Gell-Mann-Okubo mass formula with the
non-singlet masses (\ref{nonsinglet}). This field mixes with the
$SU(3)$ singlet $R_0$ via (\ref{M0corr}) in terms of two arbitrary 
constants $k_m^R, \gamma_R$ that parametrize the deviations from the
nonet limit. Although the matrix elements are of
chiral order $p^2$ the matrix (\ref{M0corr}) is therefore effectively
of a very general form. In a different notation, it has been used
since the early days of resonance physics (see also
Ref.~\cite{black99}).  

The non-singlet masses (\ref{nonsinglet}) and the mass matrix 
(\ref{M0corr}) imply the inequalities
\begin{equation} 
M_L^2 ~\le ~4 M_{I=1/2}^2/3 -  M_{I=1}^2/3 ~\le ~M_H^2~.
\label{inequ}
\end{equation} 
For given nonet masses, this is a necessary and sufficient condition
for the existence of solutions in terms of parameters $M_R$, $e_m^R$,
$k_m^R$ and $\gamma_R$. In fact, there are exactly two solutions for a
given set of masses satisfying (\ref{inequ}) that differ only in the 
sign of the mixing angle
$\theta$ and in the value of $k_m^R$. The two solutions are physically
inequivalent but we discuss in the following only the solutions with 
$\theta \ge 0$ that are closer to the nonet limit.

Because of the new terms in (\ref{Lnl}) 
there are now additional contributions of both $O(p^4)$ and
$O(p^6)$ to the effective Lagrangian (\ref{Seff}) from resonance exchange:
\begin{eqnarray}  
\cL_{\rm eff}^{\rm add} &=& -~ \displaystyle\frac{\gamma_R}{6
  M_R^2(1+\gamma_R)} \langle g_2^R\rangle^2  \nl
&& +~ \displaystyle\frac{k_m^R}{\sqrt{3} M_R^4(1+\gamma_R)}
\langle g_2^R \rangle \left(\langle g_2^R \chi_+ \rangle -
\langle g_2^R \rangle \langle \chi_+ \rangle /3 \right) \nl
&& -~ \displaystyle\frac{\gamma_R(2+\gamma_R)}{6 M_R^4(1+\gamma_R)^2}
\langle \nabla g_2^R\rangle^2 
- \displaystyle\frac{\gamma_R}{3 M_R^2(1+\gamma_R)}
\langle g_2^R\rangle  \langle g_4^R \rangle \nl
&& - ~\displaystyle\frac{2 \gamma_R}{3 M_R^4(1+\gamma_R)} \langle
  g_2^R\rangle \langle g_2^R h_2^R \rangle 
+ \displaystyle\frac{\gamma_R^2}{9 M_R^4(1+\gamma_R)^2}
\langle g_2^R\rangle^2  \langle h_2^R\rangle ~.\label{Sadd}
\end{eqnarray} 
Anticipating possible large values of the parameter $\gamma_R$, we have
written down the full expressions instead of expanding in $\gamma_R$.
We recall that $\gamma_R$ is of
zeroth order in the chiral expansion, albeit sub-leading in $1/N_c$.

\paragraph{5.}
Before turning to our main subject of scalar mesons, we briefly review
the status of the other low-lying meson resonances on the basis of the
general mass Lagrangian (\ref{Lnl}) with isosinglet mass matrix 
(\ref{M0corr}).
\paragraph{$\mbf{1^{--}}$} \mbox{   } \\[.2cm] 
The lowest-lying vector meson nonet consists of $\rho(770)$,
$\omega(782)$, $K^*(892)$ and $\phi(1020)$. From the masses in 
Ref.~\cite{rpp02} one obtains the parameters and the mixing angle
collected in Table \ref{tab1}. Not surprisingly, the vector mesons
make up an almost ideal nonet.
\paragraph{$\mbf{2^{++}}$} \mbox{   } \\[.2cm] 
The lightest nonet of tensor mesons consists of $f_2(1270)$,
$a_2(1320)$, $K_2^*(1430)$ and $f_2^\prime(1525)$. The alternative
singlet candidate $f_2(1430)$ (omitted from the summary table of
Ref.~\cite{rpp02}) does not satisfy the inequalities (\ref{inequ}).
The corresponding parameters and mixing angle in Table \ref{tab1}
document the well-known fact that also the tensor mesons are close to
an ideally mixed nonet.
\paragraph{$\mbf{1^{++}}$} \mbox{   } \\[.2cm] 
The unambiguous states in this nonet are $a_1(1260)$, $f_1(1285)$ and 
$f_1(1420)$. The strange isodoublet partner could be $K_1(1270)$ 
or $K_1(1400)$ or a mixture of these two states \cite{suzuki}.
Without mixing, only the $K_1(1270)$ satisfies the inequalities
(\ref{inequ}). The ALEPH data for $\tau \to K_1 \nu_\tau$
\cite{barate} are also consistent with a dominant $^3P_1$ nature of 
$K_1(1270)$. Neglecting a possible isodoublet mixing, the masses for
the $1^{++}$ nonet give rise to the solution in Table \ref{tab1}
implying a substantial deviation from ideal mixing (see also 
Ref.~\cite{ck97}).
\paragraph{$\mbf{1^{+-}}$} \mbox{   } \\[.2cm] 
The unambiguous states of this nonet are $h_1(1170)$ and $b_1(1235)$. 
Consistent with the assignment of $K_1(1270)$  to the $1^{++}$ nonet,
the strange member of the $1^{+-}$ nonet must be $K_1(1400)$. As
before, the situation could be more involved due to
isodoublet mixing. For the final isosinglet member of the nonet, the
two candidates are $h_1(1380)$ and $h_1(1595)$, neither of which enjoys 
the status of being listed in the PDG summary table \cite{rpp02}. With 
$K_1(1400)$ the strange state in this nonet, there is a clear
preference: only $h_1(1595)$ satisfies the inequalities 
(\ref{inequ}). The resulting solution can be found in Table
\ref{tab1}. It implies an almost ideal mixing although the sub-leading
parameter $k_m^R$ is not very small in this case.
\paragraph{$\mbf{0^{-+}}$} \mbox{   } \\[.2cm]
The well-established states are $\pi(1300)$ and $\eta(1295)$. The
$K(1460)$ does not appear in the PDG summary table but it is listed in
the full review with a mass of either 1400 or 1460 MeV. Again, the 
inequalities (\ref{inequ}) may serve as a guide. Only the lower mass of 1400
MeV allows for the inclusion of the heavy isoscalar $\eta(1440)$ with a
mass of at least 1430 MeV. There is growing evidence that there are
actually two different $0^{-+}$ states in that region \cite{rpp02}
and we therefore
take $M_{\eta(1440)}=1470$ MeV. The solution with positive $\theta$ is
again close to ideal mixing as shown in Table \ref{tab1}.

Summarizing the situation for the $1^{--}$, $2^{++}$, $1^{++}$,
$1^{+-}$ and $0^{-+}$ nonets, the sub-leading parameter $k_m^R$ is in
all cases 
substantially smaller in magnitude than $e_m^R$. With the possible
exception of $1^{++}$ that may be subject to isodoublet mixing with
$1^{+-}$, the singlet-octet mixing is close to ideal. 
All five multiplets display the standard hierarchy: $e_m^R < 0$ in 
all cases.

\renewcommand{\arraystretch}{1.2}
\begin{table}[ht]
\begin{center}
\begin{tabular}{|l|c|cccc|}
\hline
 & & & & & \\
& \hspace*{.1cm} $\theta$(degrees) \hspace*{.1cm} & \hspace*{.2cm} $M_R$(GeV) 
\hspace*{.2cm} & \hspace*{.5cm} $e_m^R$ 
\hspace*{.5cm} & \hspace*{.5cm} $k_m^R$ \hspace*{.5cm} & 
\hspace*{.5cm} $\gamma_R$ \hspace*{.5cm} \\
 & & & & & \\ \hline  
$1^{--}$ & 39 & 0.760 & - 0.23 & - 0.02 & 0.09 \\
$2^{++}$ & 32 & 1.309 & - 0.34 & - 0.03 & - 0.08 \\
$1^{++}$ & 79 & 1.226 & - 0.12 & 0.04 & 0.29 \\
$1^{+-}$ & 37 & 1.215 & - 0.50 & - 0.18 & - 0.01 \\
$0^{-+}$ & 28 & 1.292 & - 0.30 & 0.07 & - 0.05 \\
\hline 
\end{tabular}
\end{center}
\caption{Singlet-octet mixing angle $\theta$ and 
  parameters $M_R$, $e_m^R$, $k_m^R$ and $\gamma_R$ of the mass Lagrangian
  (\ref{Lnl}) for all light meson nonets except the scalars. The input 
  masses are taken from Ref.~\cite{rpp02}.}
\label{tab1} 
\end{table}

\paragraph{6.} 
Let us now focus on the scalar mesons.  Within the framework set up in 
the previous sections, we want to identify those states which, in the
large-$N_c$ limit, make up the lowest-lying nonet of scalar resonances
($0^{++}$).  This is not a straightforward task because $1/N_c$
corrections are known to significantly affect the dynamics of the
scalar sector.  In particular, in this sector the spectrum of 
QCD$_{\infty}$ seems to differ from the spectrum of QCD in the
following sense \cite{oo99,jop00}: the
inclusion of sub-leading effects in $1/N_c$ in the theoretical description
of physical processes (e.g., via loops and unitarization) generates
poles in the S-matrix that have no correspondence to the original
mass parameters of the effective Lagrangian. This leads to the notion
of ``pre-existing'' and ``dynamically generated'' resonances
\cite{oo99,tr96}.

This general feature can be understood within the analysis of
Refs.~\cite{oo99,jop00} for pseudoscalar meson meson scattering.
In the large-$N_c$ limit, the amplitudes are described by tree-level
exchange of Goldstone modes and lowest-lying resonances, as described
by CHPT and the chiral invariant effective Lagrangian 
(\ref{SLagr4}). $1/N_c$ corrections are introduced by chiral loops
and a suitable unitarization procedure (like N/D or the inverse
amplitude method).  As a general result, one finds that the full S-wave
amplitudes display not only
``pre-existing'' poles (associated with the mass parameters appearing
in the chiral resonance Lagrangian), but also ``dynamically
generated'' poles appearing as an effect of the strong S-wave
interaction.  The $\sigma(600)$ (see also Ref.~\cite{cgl01} where this
state emerges in an analysis of the Roy equations for $\pi\pi$
scattering) and $\kappa(900)$ are examples of such ``dynamically 
generated'' poles.  According to Ref.~\cite{oo99}, the $a_0(980)$ 
falls in this category as well.

The ``dynamically generated'' poles decouple in the limit of large
$N_c$. Only the ``pre-existing'' scalar states survive in 
this limit. We assume then that the latter can be
described with a chiral resonance Lagrangian to
understand the mass spectrum and the gross
features of the $S \rightarrow P_1 \, P_2$ couplings, 
with the explicit realization \cite{egpr89}
\begin{equation} 
g_2^{S} = c_d u_\mu u^\mu + c_m \chi_+ 
\label{eq:cdcm}
\end{equation} 
in the Lagrangian (\ref{SLagr4}).
We report here, for future reference, the two-pseudoscalar meson
couplings.  For the non-singlet scalar fields one has (only the
positively charged scalar fields are displayed for simplicity)
\begin{eqnarray} 
\cL (S^+_{I=1}, S^+_{I=1/2} \to 2 \rm{~mesons}) &=& \nl
 & & \hspace*{-3.5cm} S^+_{I=1} \displaystyle\frac{2}{F^2}
 \cdot \left\{ c_d \left(\displaystyle\frac{2}{\sqrt{6}}
\partial_\mu \pi^- \partial^\mu \eta_8 + \partial_\mu K^- 
\partial^\mu K^0 \right)\right. \nl 
 & & \hspace*{-2cm} \left. - c_m \left(\displaystyle\frac{2 M_\pi^2}
{\sqrt{6}} \pi^- \eta_8 +  M_K^2 K^- K^0 \right) \right\} 
\label{a0kappa}  \\
& & \hspace*{-3.7cm} + S^+_{I=1/2} \displaystyle\frac{1}{F^2} \cdot 
\left\{ c_d \left(\sqrt{2} \partial_\mu K^- \partial^\mu \pi^0
+ 2 \partial_\mu \pi^- \partial^\mu \ol{K^0} 
- \displaystyle\frac{2}{\sqrt{6}}\partial^\mu K^- \partial_\mu \eta_8 
\right)\right. \nl
 & & \hspace*{-2cm} \left. - c_m \left(\displaystyle\frac{M_K^2+M_\pi^2}
{\sqrt{2}}( K^- \pi^0 + \sqrt{2} \pi^- \ol{K^0}) + 
\displaystyle\frac{3 M_\pi^2 - 5 M_K^2}{\sqrt{6}} K^- \eta_8  \right) 
\right\}  \ , \nn
\end{eqnarray}
while for the  strange and  non-strange isosinglet fields 
of Eq.~(\ref{flavour}) one finds 
\begin{eqnarray} 
\cL (S_{\rm non-strange}, S_{\rm strange} \to 2 \rm{~mesons}) &=& 
\nonumber\\*
 & & \hspace*{-7cm}  S_{\rm non-strange} \displaystyle\frac{\sqrt{2}}{F^2}
 \cdot \left\{ c_d \left(\partial_\mu \pi^0 \partial^\mu \pi^0 + 2
\partial_\mu \pi^+ \partial^\mu \pi^- + \partial_\mu K^+ \partial^\mu
K^- + \partial_\mu K^0 \partial^\mu \ol{K^0} 
+ \displaystyle\frac{1}{3}\partial_\mu \eta_8 \partial^\mu \eta_8
\right)\right. \nl 
 & & \hspace*{-3.9cm} \left. - c_m \left( M_\pi^2 
(\pi^0 \pi^0 + 2 \pi^+ \pi^-) + 
M_K^2 (K^+ K^- + K^0 \ol{K^0})+ \displaystyle\frac{M_\pi^2}{3}
\eta_8 \eta_8 \right) \right\} \nl
& & \hspace*{-7.3cm} + S_{\rm strange} \displaystyle\frac{1}{F^2} \cdot 
\left\{ c_d \left(2 \partial_\mu K^+ \partial^\mu
K^- + 2 \partial_\mu K^0 \partial^\mu \ol{K^0} 
+ \displaystyle\frac{4}{3}\partial_\mu \eta_8 \partial^\mu \eta_8
\right)\right. \nl
 & & \hspace*{-4.7cm} \left. - c_m \left(2 M_K^2 (K^+ K^- + K^0
\ol{K^0}) + \displaystyle\frac{4}{3}(2 M_K^2 - M_\pi^2)
\eta_8 \eta_8 \right) \right\}~.
\label{S0S8}  
\end{eqnarray} 
Thus, $S_{\rm strange}$ does not couple to pions as expected. On the other
hand, $S_{\rm non-strange}$ couples to both strange and non-strange mesons
with full strength. This is a straightforward consequence of (softly
broken) $SU(3)$ incorporated in the chiral expansion. At the hadronic
level, there is no fundamental difference between two- and four-quark 
states. 

The couplings $c_d$ and $c_m$ were originally fixed \cite{egpr89} by
requiring that the phenomenologically determined values for $L_{5}^r
(M_\rho)$ and $L_{8}^r (M_\rho)$ are saturated by scalar resonance
exchange. This led to $c_d \simeq 32$ MeV and $c_m \simeq 42$ MeV.
Later on, independent information on these couplings was obtained from
the study of QCD short-distance constraints on the SS correlator and on
the scalar form factor \cite{jop00,ap02}; in the single-resonance
approximation, one finds
\beq
c_d = c_m = F_\pi / 2 = 46 {\rm ~MeV} \ . 
\eeq
Finally, results consistent with the above have been obtained 
by fitting experimental meson meson phase shifts within a
chiral unitary approach \cite{oo99,jop00}, with the best fits 
pointing to somewhat smaller values ($c_d \sim c_m \sim 20$ MeV). 

We can now identify possible candidates for the lightest scalar
nonet at large $N_c$, and try to discriminate between them on a
phenomenological basis.  The $I=1/2$ member of the nonet is identified
without controversy with $K_0^{*}(1430)$. For the $I=0$ states, we
only consider $f_0(980)$ and $f_0(1500)$ as candidates, excluding
$f_0(1370)$ for the arguments given in Ref.~\cite{minkochs99}.
Two scenarios then arise, depending on the assignment for the $I=1$
state.
\begin{itemize}
\item[A:] If we assume that the $a_0(980)$ is dynamically 
generated \cite{oo99} (and makes up an octet together 
with $\sigma(600)$ and $\kappa(900)$ in the  $SU(3)$ limit), 
the remaining candidates for a nonet are 
$$ f_0 (980),   K_0^{*}(1430),  a_0 (1450),  f_0 (1500) \ .  $$
\item[B:] On the other hand, assuming that  $a_0(980)$ is a 
pre-existing state in the large-$N_c$ limit, the nonet would be 
composed by \cite{minkochs99}
$$ f_0 (980), a_0(980),  K_0^{*}(1430), f_0 (1500) \ .  $$
\end{itemize}

Contrary to what we have found for other resonance  multiplets, both 
scenarios A and B are very poorly described by the strict nonet limit
($k_m^S=\gamma_S=0$).  In scenario A, the isoscalar masses turn out to be
$M_L = 1.35$ GeV and $M_H = 1.47$ GeV, with a sizable deviation of
$M_L$ from $M_{f_0(980)}$.  Moreover, the dual ideal mixing angle would
imply $S_L= S_{\rm strange}$. Consequently, the $f_0(980)$ would not
couple to two pions, which is not phenomenologically acceptable.  In 
scenario B, the isoscalar masses turn out to be $M_L = 0.985$ GeV and 
$M_H = 1.74$ GeV, the latter being significantly bigger than $M_{f_0(1500)}$.

These observations point towards sizable nonet-breaking effects.
By fitting to the relevant mass spectra\footnote{In scenario B, using 
the input masses from Ref.~\cite{rpp02}, the inequality (\ref{inequ}) 
is violated at the upper end so that the spectrum cannot be fitted in
terms of $M_S$, $e_m^S$, $k_m^S$, and $\gamma_S$. For the numerics in Table
\ref{tab2} we use $M_{I=1/2}=1.39$ GeV, which is fully consistent with 
the PDG entry, given the large width of $294$ MeV for $K_0^*(1430)$.}
\cite{rpp02}, we obtain the parameters and the mixing angle collected
in Table \ref{tab2}. 
The non-negligible $1/N_c$ effects translate into relatively large
values for the couplings $k_m^S$ and $\gamma_S$, as compared to $e_m^S$. 
Case A displays an inverted hierarchy ($e_m^S > 0$) and a mixing angle 
close to ideal. Case B displays the standard hierarchy ($e_m^S < 0$) and 
a mixing angle close to zero, corresponding to the scenario of 
Ref.~\cite{minkochs99}.
 
\renewcommand{\arraystretch}{1.2}
\begin{table}[ht]
\begin{center}
\begin{tabular}{|l|c|cccc|}
\hline
 &  & & & & \\
& \hspace*{.1cm} $\theta$ (degrees) \hspace*{.1cm} 
& \hspace*{.2cm} $M_S$(GeV) \hspace*{.2cm} 
& \hspace*{.5cm} $e_m^S$ 
\hspace*{.5cm} & \hspace*{.5cm} $k_m^S$ \hspace*{.5cm} & 
\hspace*{.5cm} $\gamma_S$ \hspace*{.5cm} \\
 & & & & & \\ \hline  
$0^{++} (A)$ & 30 & 1.48  &  0.20  & - 1.00 & - 0.35 \\
$0^{++}$(B) & 6.8 & 0.94  & - 1.06 & 1.02 & - 0.71 \\
\hline 
\end{tabular}
\end{center}
\caption{Singlet-octet mixing angle $\theta$ and 
  the parameters $M_S$, 
  $e_m^S$, $k_m^S$, $\gamma_S$ for the $0^{++}$ light scalar nonet 
   corresponding to scenarios A and B.
  The input masses are taken from Ref.~\cite{rpp02}, with the exception of 
  scenario B, where we use $M_{I=1/2}=1.39$ GeV.}
\label{tab2} 
\end{table}

\paragraph{7.} 
In order to discriminate between the two options, we start with a
qualitative argument. Scenario A is attractive because two
full multiplets are identified, one nonet of pre-existing states
and another octet of dynamically generated poles, which decouple in
the large-$N_c$ limit. Put in another way, the role of the 
well-established $a_0(1450)$ is not clear in scenario B.
Although the near-degeneracy of $a_0(980)$ and $f_0(980)$ can be
accommodated in scenario B we are in this case very far from the nonet 
limit that would naturally explain the degeneracy.

More quantitative arguments can be given if we assume that the chiral
resonance Lagrangian reproduces at least the salient features of the 
phenomenology of $S \rightarrow P_1 \, P_2$ decays.
First of all, we determine the contributions from scalar resonance
exchange to the LECs of $O(p^4)$. From the effective Lagrangians 
(\ref{Seff}) and (\ref{Sadd}) one obtains
\begin{equation} 
\begin{array}{lll}   
L_1^S = - \displaystyle\frac{\gamma_S c_d^2}{6 M_S^2(1+\gamma_S)}~,
\hspace*{.5cm}  &
L_3^S = \displaystyle\frac{c_d^2}{2 M_S^2}~,\hspace*{.5cm} &
L_4^S = - \displaystyle\frac{\gamma_S c_d c_m}{3 M_S^2(1+\gamma_S)}~,\\
L_5^S = \displaystyle\frac{c_d c_m}{M_S^2}~, \hspace*{.5cm} & 
L_6^S = - \displaystyle\frac{\gamma_S c_m^2}{6 M_S^2(1+\gamma_S)}~, 
\hspace*{.5cm} &
L_8^S = \displaystyle\frac{c_m^2}{2 M_S^2}~.    
\end{array}
\end{equation}  
Assuming $c_m=c_d=F_\pi/2$, the numerical values of the LECS (in units
of $10^{-3}$) for the two scenarios are
\begin{equation} 
\begin{array}{ccccccc}
 & \hspace*{.3cm}  L_1^S \hspace*{.3cm} & \hspace*{.3cm} L_3^S 
\hspace*{.3cm} & \hspace*{.3cm} L_4^S \hspace*{.3cm} & \hspace*{.3cm} 
L_5^S \hspace*{.3cm} & \hspace*{.3cm}
 L_6^S \hspace*{.3cm} & \hspace*{.3cm} L_8^S \hspace*{.3cm} \\
 {\rm A} \hspace*{.6cm} & 0.1 & 0.5 & 0.2 & 1.0 & 0.1 & 0.5 \\
{\rm B} \hspace*{.6cm} & 1.0 & 1.2 & 2.0 & 2.4 & 1.0 & 1.2
\end{array}
\end{equation}
Although some of the LECs (especially $L_5$ and $L_8$ \cite{abt01})
may have to be reanalysed on the basis of our work there is a certain
preference for scenario A. The main drawback of scenario B is the
large value of $L_4^S$, even for smaller values of $c_d,c_m$. 
 
One can check that the resonance Lagrangian (\ref{a0kappa})
works reasonably well for 
the decays of the $I=1$ and $I=1/2$ states, especially when
considering ratios of decay widths\footnote{In the case of $c_m=c_d$, 
all rates are proportional to $c_d^2$, which cancels in the ratios.}.  
Turning to the isoscalar sector, both scenarios A and B correctly
predict that $f_0(980)$ couples predominantly to the $\pi \pi$ state.
On the other hand, there is a marked difference between the two
scenarios for $f_0(1500)$ due to very different mixing angles. Within 
option A, the two-pion mode is severely suppressed because of the
nearly ideal mixing angle. In scenario B, on
the other hand, $f_0(1500)$ couples strongly to two pions. Although
for a state as heavy as the $f_0(1500)$ the relative branching ratios 
cannot be determined reliably from the tree-level amplitude only,
the relatively small total width $\Gamma[f_0(1500)]=109$ MeV \cite{rpp02}
seems to favour again scenario A because the $\pi\pi$ channel is
strongly suppressed in this case.

Finally, the contributions of scalar resonances to the LECs of
$O(p^6)$ are also quite different in the two cases, mainly
because they scale as $1/M_S^4$. Moreover, due to the smaller value of 
$e_m^S$, scenario A leads to 
\begin{itemize} 
\item better behaved corrections of $O(p^6)$ to masses
and decay constants of pseudoscalar mesons \cite{abt00}; 
\item  more reasonable contributions  to isospin breaking effects 
in $ K \rightarrow \pi \pi $ decays \cite{gv99,cenp032}.
\end{itemize}

\paragraph{8.}
The scalar mesons are very likely the only light meson resonances where
large $N_c$ together with chiral symmetry fails dramatically. We have
investigated a scenario where the deviations from nonet symmetry
occur only in the bilinear part of the resonance Lagrangian and
therefore in the scalar mass matrix. 

We have identified two possible scenarios for the lightest scalar
nonet that survives in the large-$N_c$ limit, and we have discussed
arguments to discriminate between the two.  Analysis of the
mixing of isoscalars seems to favour an inverted hierarchy for the
scalar mesons where the isotriplet states $a_0(1450)$ are heavier than
the strange particles $K_0^*(1430)$ (Scenario A).
The main features of the decays of scalar resonances to two
pseudoscalars can also be understood within this framework
although a more detailed dynamical study would be needed for a
quantitative description. Altogether, our analysis favours a
lightest ``pre-existing'' scalar nonet consisting of the states
$f_0(980), K_0^{*}(1430), a_0(1450), f_0(1500)$.

Besides providing arguments for the composition of the lightest scalar
nonet surviving in the large-$N_c$ limit, our analysis has
implications for all
observables that are especially sensitive to scalar resonance exchange. 
Among those are the pseudoscalar meson masses and decay
constants \cite{abt00}. Another important application is
isospin violation in the CP-violating parameter $\epsp$ \cite{gv99}.
Our results imply that the coupling constant $e_m^S$ (first considered
in Ref.~\cite{abt00}) and the nonet-breaking coupling $k_m^S$ in the 
Lagrangian (\ref{Lnl}) produce
isospin-violating contributions of similar size. The implications for 
$\epsp$ will be considered elsewhere \cite{cenp032}.

\vfill
\paragraph{Acknowledgements}
\noindent 
We would like to thank J. Portoles and J.J. Sanz Cillero for helpful
discussions. We are also grateful to J. Bijnens, J. Gasser and P. 
Minkowski for useful comments and suggestions.
The work of V.C. and A.P. has been supported in part by
MCYT, Spain (Grant No. FPA-2001-3031) and by ERDF funds from the
European Commission.

 

\end{document}